\documentclass[preprint,12pt]{elsarticle}
\usepackage{subfigure}
\usepackage{amsmath}
\usepackage{amssymb}
\usepackage{url}
\biboptions{sort&compress}
\usepackage[a4paper,total={6in, 8in}]{geometry}
\usepackage{atbegshi}% http://ctan.org/pkg/atbegshi
\AtBeginDocument{\AtBeginShipoutNext{\AtBeginShipoutDiscard}}

\journal{Physics Letters A}

\begin{document}

\begin{frontmatter}

%opening
\title{Crossover from BKT to first-order transition induced by higher-order terms in 2D $XY$ models}
\author{Milan \v{Z}ukovi\v{c}}
\ead{milan.zukovic@upjs.sk}
\address{Department of Theoretical Physics and Astrophysics, Institute of Physics, Faculty of Science, Pavol Jozef \v{S}af\'arik University in Ko\v{s}ice, Park Angelinum 9, 041 54 Ko\v{s}ice, Slovakia}

\maketitle

\begin{abstract}
We study phase transitions in $XY$ models, generalized by inclusion of $n$ higher-order pairwise interactions of equal strength, by Monte Carlo simulation. We inspect the evolution of the behavior at the transition point in the models with gradually increasing number of terms. It is found that by adding new terms the signatures of the Berezinskii-Kosterlitz-Thouless (BKT) transition, observed in the standard $XY$ ($n=1$) model, gradually change to those characteristic for a first-order phase transition. By a finite-size scaling analysis we determine the critical number of the terms for which the first-order transition appears as $n_c=6$. It is also found that a pseudo-first-order transition is present for a smaller value of $n=5$. However, this transition becomes true first-order if the couplings at the respective terms are allowed to increase, thus bringing the critical number down to $n_c=5$. In general, a more rapid increase of the coupling intensity appears to support the first-order transition, however, a too fast increase may result in the splitting of the single transition to multiple transitions. Consequently, the minimal number of the terms required for the change of the BKT phase transition to first order in the present model with arbitrary couplings is estimated to be $2 < n_c \leq 5$.

\end{abstract}


%\PACS 05.50.+q \sep 64.60.De \sep 75.10.Hk \sep 75.30.Kz \sep 75.50.Ee \sep 75.50.Lk

\end{frontmatter}

\section{Introduction}

It is known that in two-dimensional (2D) spin systems with short-range interactions no continuous symmetry breaking that would result in a long-range ordering (LRO) can occur~\cite{merm66}. Nevertheless, a quasi-LRO is possible due to the presence of topological excitations. In particular, the simple 2D $XY$ with model two spin components (or planar rotator), with the Hamiltonian ${\mathcal H_1}=-J_1\sum_{\langle i,j \rangle}\cos(\phi_{i,j})$, where the interaction $J_1>0$ is limited to nearest-neighbor pairs forming the angle $\phi_{i,j}=\phi_{i}-\phi_{j}$, is well known to show a topological Berezinskii-Kosterlitz-Thouless (BKT) phase transition~\cite{berez71,kt73}. The topological defects, called vortices and antivortices, are free at higher temperatures but at the BKT transition temperature $T_{\rm BKT}$ they bind in pairs. This results in the emergence of the quasi-LRO BKT phase (critical for all temperatures $T \leq T_{\rm BKT}$), which is characterized by an algebraically decaying correlation function.

Various generalizations of the $XY$ model have been proposed to explain novel phenomena, observed in realistic systems. Perhaps, the most studied generalization assumes the presence of the $q$-order (pseudo)nematic term, with the Hamiltonian ${\mathcal H_q}=-J_q\sum_{\langle i,j \rangle}\cos(q\phi_{i,j})$, where $q=2,3,\hdots$. The resulting generalized $J_1-J_q$ model with the Hamiltonian ${\mathcal H} = {\mathcal H_1} + {\mathcal H_q}$ was extensively studied for the $q=2$ case in effort to explain phases and phase transitions in different experimental realizations, such as liquid crystals~\cite{lee85,geng09}, superfluid A phase of $^3{\rm He}$~\cite{kors85}, and high-temperature cuprate superconductors~\cite{hlub08}. The phase diagram of such a model features a nematic quasi-LRO phase, which is separated from the magnetic one at lower temperatures by the phase boundary belonging to the Ising universality class~\cite{lee85,kors85,carp89}.

Surprisingly, the increasing order of the nematic term has been found to lead to a dramatic change of the critical behavior. In particular, the $XY$ models with $J_1-J_q$ interactions for $q \geq 4$ were demonstrated to display four possible ordered phases with the phase boundaries belonging to various (Potts, Ising, or BKT) universality classes~\cite{pode11,cano14,cano16,nui23}. Consequently, such models could show up to three phase transitions upon their cooling from the paramagnetic phase.

Another generalization of the $XY$ model with the Hamiltonian in the form ${\mathcal H}=2J\sum_{\langle i,j \rangle}(1-[\cos^2(\phi_{i,j}/2)]^{p^2})$, in which the potential shape is controlled by the parameter $p^2$, predicted the existence of only one phase transition the nature of which, however, changes to the first order for sufficiently large values of $p^2$~\cite{doma84,himb84,sinh10a,sinh10b}. The crossover to the first-order transition has also been suggested for a generalized 2D $XY$ model in which spins are allowed to have three components. In particular, the model with the Hamiltonian ${\mathcal H}=-J\sum_{\langle i,j \rangle}(\sin \theta_i \sin \theta_j)^q \cos(\phi_{i}-\phi_{j})$,  where $\theta$ is the azimuthal and $\phi$ polar angles, has been claimed to display the first-order transition for sufficiently large values of the generalization parameter $q$~\cite{mol06,ente06,dasilva24}, albeit this scenario was questioned in Ref.~\cite{komura11}. Nevertheless, the possibility of such a change of the transition order in the SO($n$) invariant $n$-vector models with a sufficiently nonlinear potential shape (showing a deep and narrow well) has been rigorously proved~\cite{ente02,ente05}.

The $XY$ model, generalized by inclusion of up to an infinite number of higher-order pairwise interactions with an exponentially decreasing strength has also been found to show a crossover to the first-order transition regime for the cases with highly nonlinear shape of the potential well~\cite{zuko17}. In this model it was the case for either a very large number of the higher-order terms or a very slow decay of their strength. In the present study we consider a related model with $n$ higher-order (pseudo)nematic terms, ${\mathcal H_k}=-J_k\sum_{\langle i,j \rangle}\cos(k\phi_{i,j})$ for $k=1,\hdots,n$, of equal strength, i.e., $J_k \equiv const$. Based on the critical behavior of the model considered in Ref.~\cite{zuko17} it is expected that such conditions would maximize the degree of nonlinearity of shape of the potential well for a given number of the terms with non-increasing strength. Then, our goal is to determine the minimal number of the terms $n_c$ necessary for the change of the nature of the transition to first order.

\section{Model and Method}
\label{sec:model}
We consider the generalized $XY$ model described by the Hamiltonian
\begin{equation}
\label{Hamiltonian1}
{\mathcal H(n)}=\sum_{\langle i,j \rangle}H_{i,j}(n),
\end{equation}
where the summation runs over all nearest-neighbor pairs of spins on the lattice, $H_{i,j}(n)=-\sum_{k=1}^{n}J_k\cos(k\phi_{i,j})$ is the pairwise potential with the summation running over all $n$ terms in the Hamiltonian, $\phi_{i,j}=\phi_{i}-\phi_{j}$ is an angle between the neighboring spins $i$ and $j$ and the coupling constants are set to $J_k \equiv J=1$.  

\begin{figure}[t!]
\centering
\subfigure{\includegraphics[scale=0.5,clip]{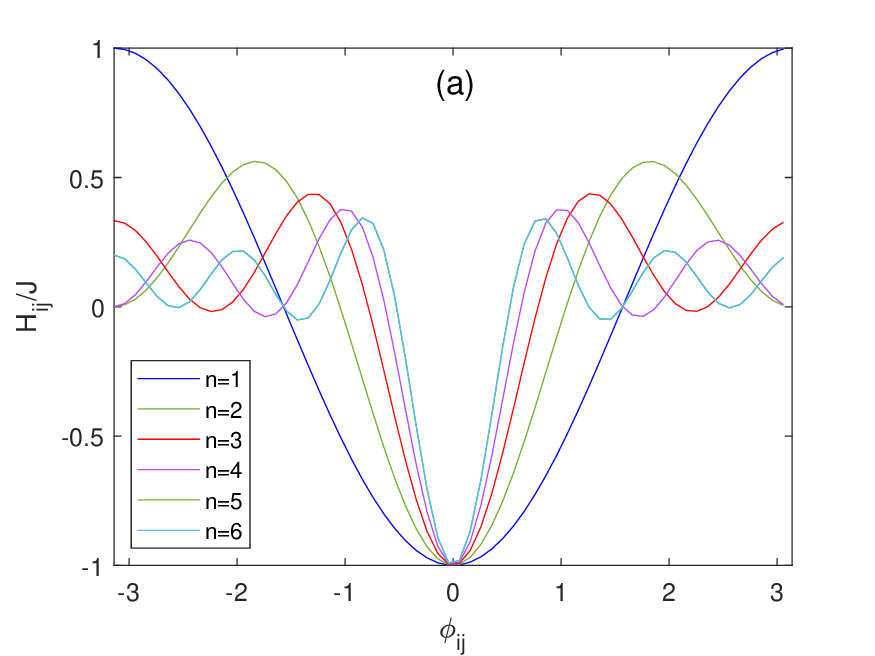}\label{fig:well_n1-6}} 
\subfigure{\includegraphics[scale=0.5,clip]{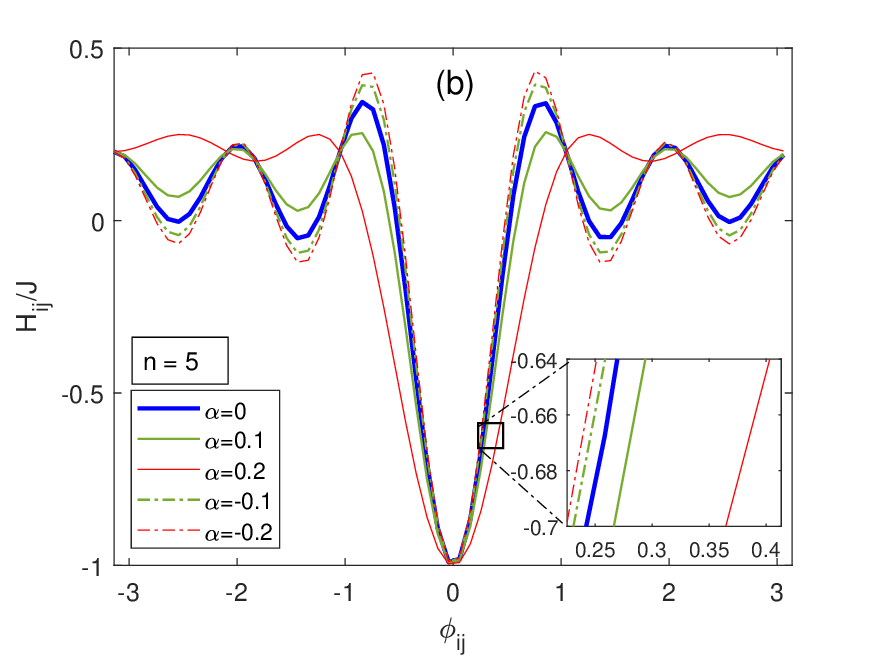}\label{fig:well_n5}} 
\caption{Pairwise potential $H_{i,j}/J$, where $J=\sum_{k=1}^{n} J_k$, for (a) $J_k =1$ and different values of $n$ and (b) $n=5$ and $J_k=1-\alpha k$ with different values of $\alpha$.}
\label{fig:well}
\end{figure}

By inspecting the pairwise potential $H_{i,j}(n)$ one can notice that, in line with our expectations, with the increasing number of the terms its shape becomes gradually narrower (Fig.~\ref{fig:well_n1-6}). Consequently, we expect that for a certain value $n_c$ the potential well becomes sufficiently narrow to trigger the crossover to the first-order transition regime~\cite{ente02,ente05}. It is supposed that the extreme narrowness of the well suppresses formation of defect pairs at low temperatures and thus facilitates their dramatic proliferation at the first-order transition point~\cite{himb84}. Figure~\ref{fig:well_n5} demonstrates the effect of the coupling constants $J_k$ on the shape of the potential well for a fixed number of the terms. In particular, instead of keeping $J_k$ constant we tentatively considered a simple linear dependence $J_k=1-\alpha k$. The case of $\alpha=0$ corresponds to the condition $J_k =1$ in our model, while $\alpha>0$ ($\alpha<0$) means decreasing (increasing) dependence. The plots support the assumption that the potential in our model corresponds to the narrowest shape for a given number of the terms ($n=5$) with non-increasing strength. One can also notice that the potential well could become even narrower if one considered the increasing dependence of $J_k$. However, as will be discussed below, such conditions can also result in splitting of the transition into two or more separate transitions instead of one transition becoming first order.

To study thermal behavior, we calculate the following quantities: the internal energy per spin $e=\langle {\mathcal H} \rangle/L^2$, the specific heat per spin
\begin{equation}
c=\frac{\langle {\mathcal H}^{2} \rangle - \langle {\mathcal H} \rangle^{2}}{L^2T^{2}},
\label{c}
\end{equation}
the magnetization
\begin{equation}
m=\langle M \rangle/L^2=\left\langle\Big|\sum_{j}\exp(i\phi_j)\Big|\right\rangle/L^2,
\label{oq}
\end{equation}
the magnetic susceptibility
\begin{equation}
\label{chi_oq}\chi = \frac{\langle M^{2} \rangle - \langle M \rangle^{2}}{L^2T},
\end{equation}
and the fourth-order magnetic Binder cumulant
\begin{equation}
\label{U}U = 1-\frac{\langle M^{4}\rangle}{3\langle M^{2}\rangle^{2}},
\end{equation}
where $\langle \hdots \rangle$ denotes thermal average.

Another important quantity is the vortex density, denoted as $\rho$. A vortex (or antivortex) is a topological defect characterized by a $2\pi$ ($-2\pi$) change in the spin angle when moving around a closed contour enclosing the core of the excitation. The vortex density can be directly computed from MC states by summing the angles between adjacent spins on each square plaquette for each equilibrium configuration. In this context, a $2\pi$ corresponds to a vortex, $-2\pi$ to an antivortex, and a value of $0$ indicates no topological defect. Finally, the vortex density $\rho$ is obtained by calculating the normalized thermodynamic average of the absolute vorticity (accounting for both vortices and antivortices), summed over the entire lattice and normalized by the system's volume.

At the standard BKT transition the internal energy and the magnetization are expected to vary smoothly, while the magnetic susceptibility will diverge as power law, characterized by the exponent $2-\eta=7/4$. The latter can be estimated by a finite-size scaling (FSS) relation
\begin{equation}
\label{xi_FSS}
\chi(L) \propto L^{2-\eta}.
\end{equation}
On the other hand, if the transition is of first order, then the internal energy and the magnetization will show in thermodynamic limit a discontinuous behavior, the thermodynamic functions like the susceptibility are supposed to scale with volume, i.e., $\chi(L) \propto L^{2}$, and the Binder cumulant is expected to plunge to negative values~\cite{tsai98}.

\section{Results}
\label{sec:results}

\begin{figure}[t!]
\centering
\subfigure{\includegraphics[scale=0.5,clip]{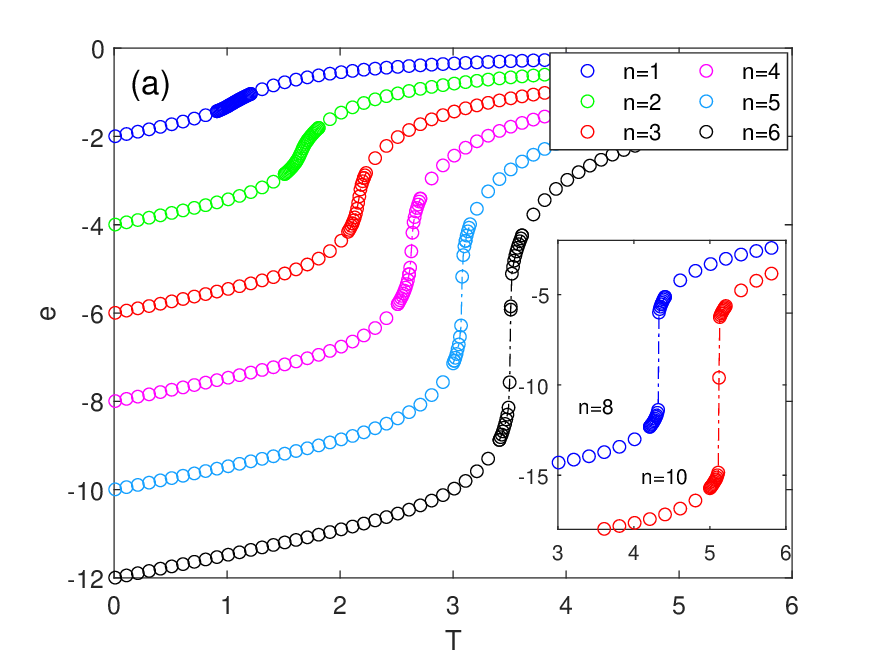}\label{fig:all_nk_e-T}} 
\subfigure{\includegraphics[scale=0.5,clip]{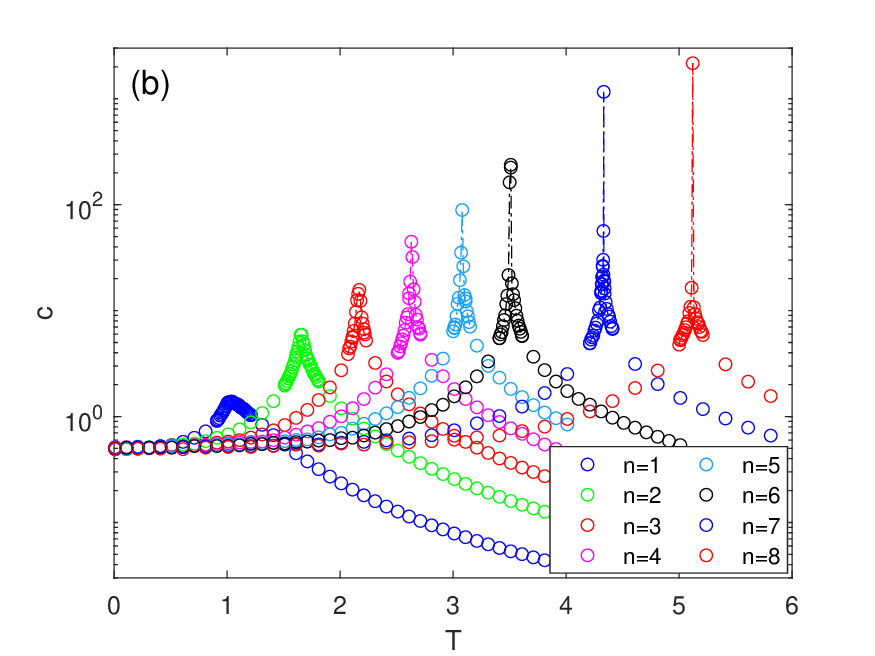}\label{fig:all_nk_c-T}} \\
\subfigure{\includegraphics[scale=0.5,clip]{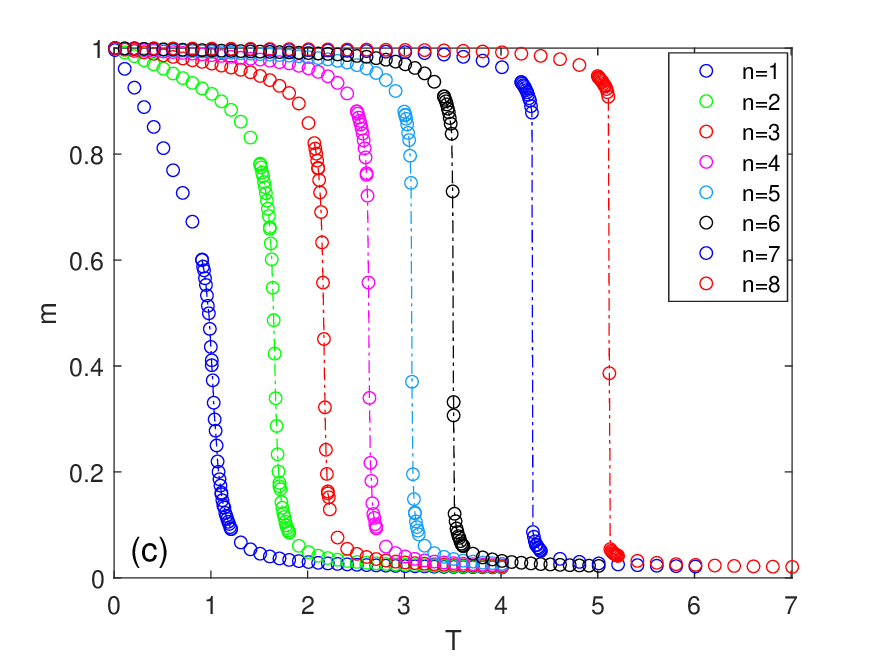}\label{fig:all_nk_m-T}}
\subfigure{\includegraphics[scale=0.5,clip]{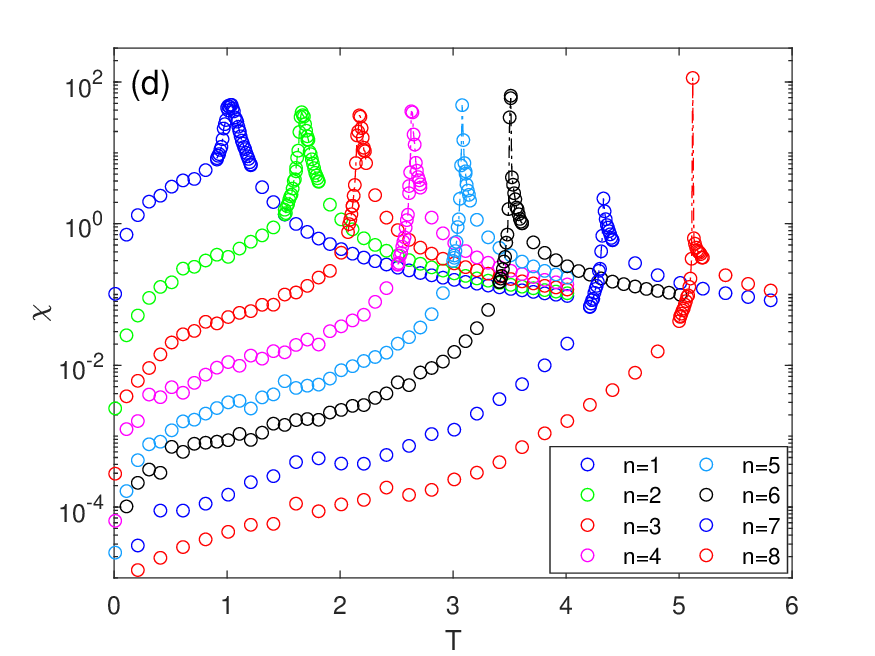}\label{fig:all_nk_chi-T}} \\
\subfigure{\includegraphics[scale=0.5,clip]{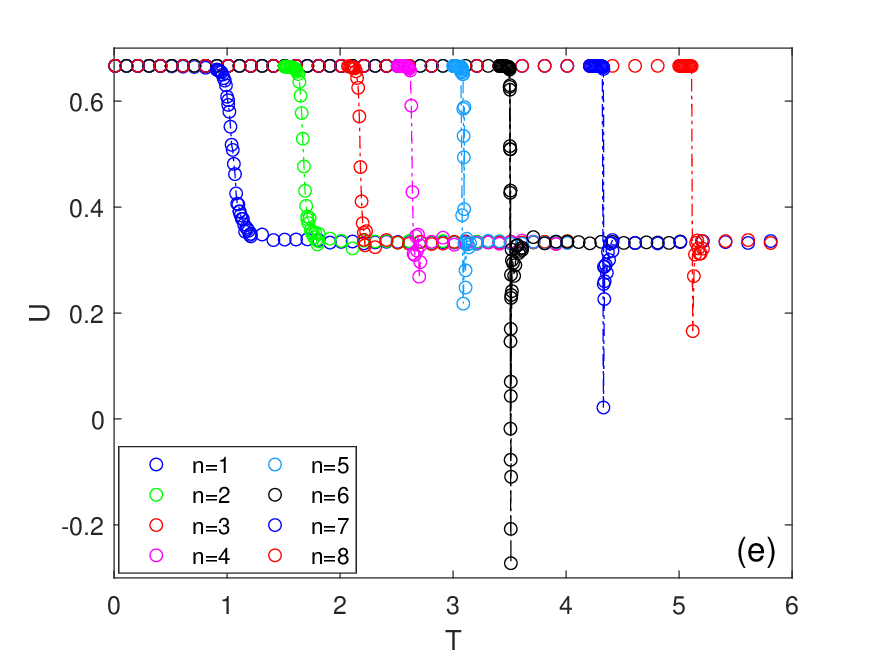}\label{fig:all_nk_u-T}}
\subfigure{\includegraphics[scale=0.5,clip]{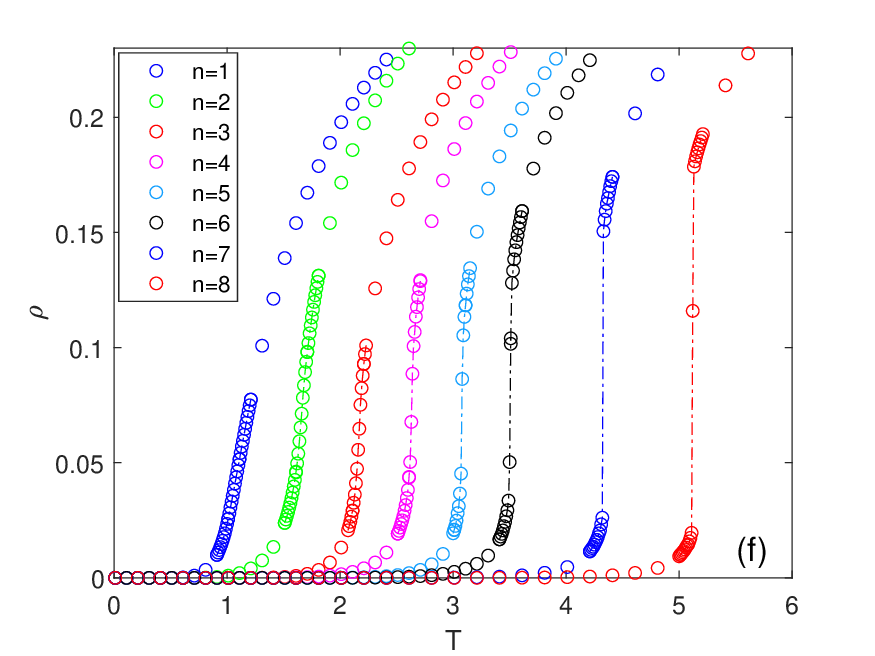}\label{fig:all_nk_rho-T}}
\caption{Temperature dependencies of (a) the internal energy, (b) the specific heat, (c) the magnetization, (d) the magnetic susceptibilities, (e) the Binder cumulant, and (f) the vortex density, for $L=60$ and different values of $n$.}
\label{fig:x-T}
\end{figure}

A rough picture about the number and nature of the phase transitions that occur in the model can be obtained by plotting temperature dependencies of the measured quantities. In Fig.~\ref{fig:x-T} such plots are presented for a gradually increasing number of the higher-order terms in the Hamiltonian. The internal energy and magnetization curves show the expected anomalies at a single transition point with apparently continuous variation for smaller values of $n$. However, with the increasing $n$ the curves at the transition point become steeper and for $n \gtrsim 6$ their behavior becomes apparently discontinuous. The corresponding response functions, i.e., the specific heat and the magnetic susceptibility, at the transition display a peak that appears rounder for smaller $n$ but becomes spike-like for $n \gtrsim 6$. The Binder cumulant with the increasing $n$ starts displaying a sharp dip, just after the transition point, which for $n \gtrsim 6$ drops to negative values\footnote{For $n=7$ and $8$ the Binder cumulant curves do not seem to reach negative values but this is only due to the limited resolution of the temperature scale and extreme narrowness of the dip. For similar reasons there is an absence of a spike at the transition in the susceptibility curve for $n=7$.}. Finally, continuous behavior for smaller $n$ that becomes discontinuous for $n \gtrsim 6$ can also be observed in the vortex density plots.

\begin{figure}[t!]
\centering
\subfigure{\includegraphics[scale=0.5,clip]{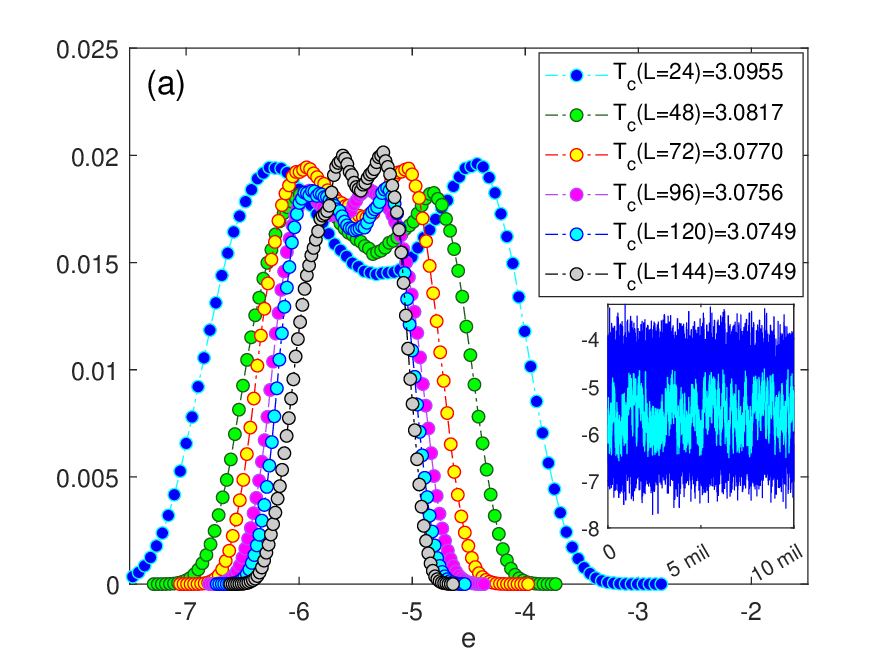}\label{fig:nk5_ene_hist}}
\subfigure{\includegraphics[scale=0.5,clip]{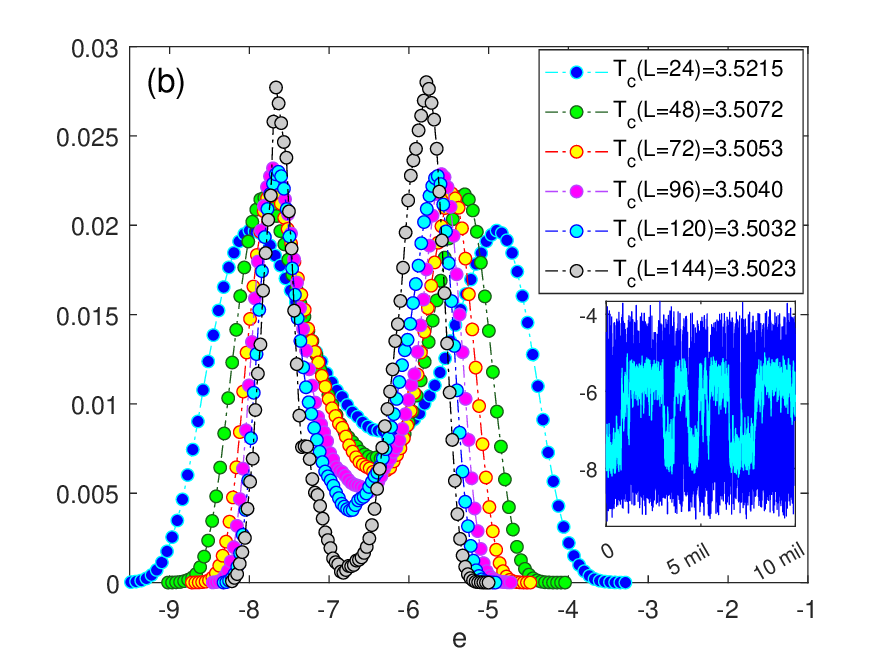}\label{fig:nk6_ene_hist}} \\
\subfigure{\includegraphics[scale=0.5,clip]{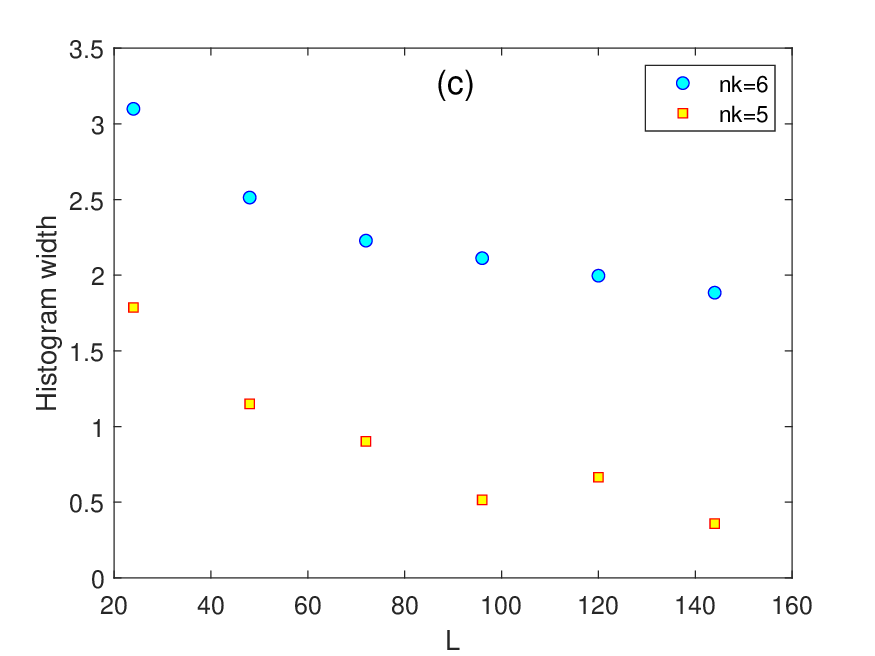}\label{fig:hist_width}}
\subfigure{\includegraphics[scale=0.5,clip]{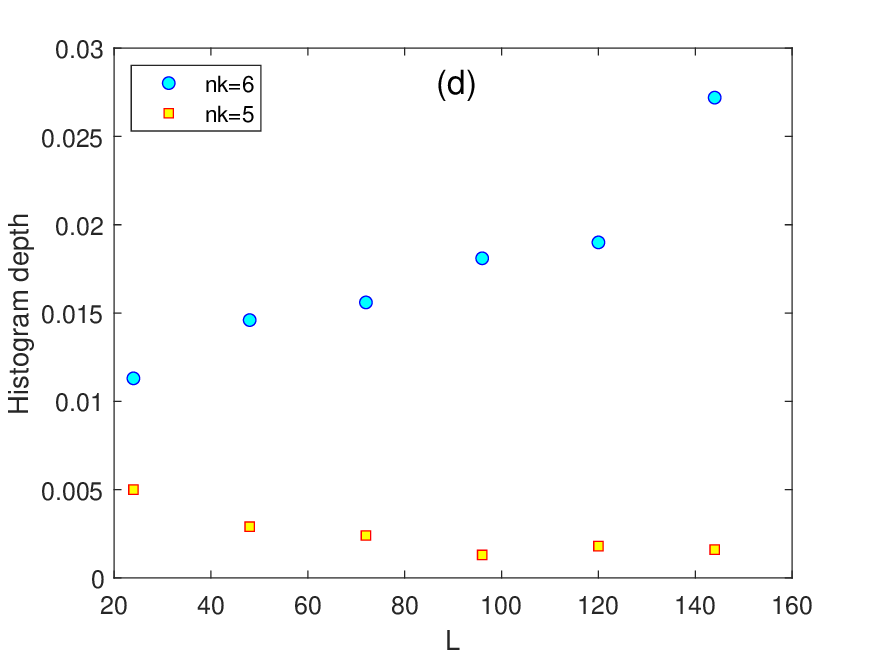}\label{fig:hist_depth}}
\caption{Top row: energy histograms at the (pseudo)transition temperatures for different lattice sizes and (a) $n=5$ and (b) $n=6$. Bottom row: (c) the widths and (d) the depths of the energy histograms at the (pseudo)transition temperatures for different lattice sizes and $n=5$ (squares) and $n=6$ (circles). The insets in (a) and (b) demonstrate the time evolution of the energy for $L=24$ (dark blue) and $L=120$ (cyan).}
\label{fig:hist}
\end{figure}

The above results indicate the existence of the crossover to the first-order transition at $n_c \approx 6$. To determine the exact value with higher reliability we perform a FSS analysis for the candidate values $n=5$ and $6$. The first-order transition features can be revealed by studying energy distributions at the transition temperature. The first-order transition is characterized by a bimodal distribution and with the increasing $L$ the heights of the peaks are expected to increase and the valley between them decrease. In the limit of $L \to \infty$ the valley should tend to zero and the distance between the peaks should approach a finite value, corresponding to the latent heat released/absorbed at the transition. As shown in Fig.~\ref{fig:hist}, the bimodal character of the energy distribution can be observed both for $n=5$ (Fig.~\ref{fig:nk5_ene_hist}) and $n=6$ (Fig.~\ref{fig:nk6_ene_hist}). Nevertheless, there is an important difference between these two cases in the behavior with the increasing $L$. While for $n=6$ the distributions display the typical first-order-transition behavior described above, for $n=5$ the behavior is quite different. Namely, with the increasing lattice size the height of the peaks does not systematically change, the valley between them does not get deeper but rather shallower and as the peaks continue to move towards each other it also becomes narrower. This is apparent from Figs.~\ref{fig:hist_width} and~\ref{fig:hist_depth}, in which the energy histogram width and depth are plotted as functions of $L$ for both cases of $n=5$ and $n=6$. The difference between these two cases are also manifested in the insets next to the histograms that show the energy time series for a small ($L=24$) and larger ($L=120$) system sizes. Thus we believe that the observed double-peak structure for $n=5$ is just a finite-size effect and in the thermodynamic limit it will vanish. This behavior is typical for a so-called pseudo-first-order transition that has also been observed in some other systems, such as the $XY$ model generalized by inclusion of higher-order terms with an exponentially decreasing strength~\cite{zuko17} or 4-state Potts and $J_1-J_2$ Ising models~\cite{jin12}.

\begin{figure}[t!]
\centering
\includegraphics[scale=0.5,clip]{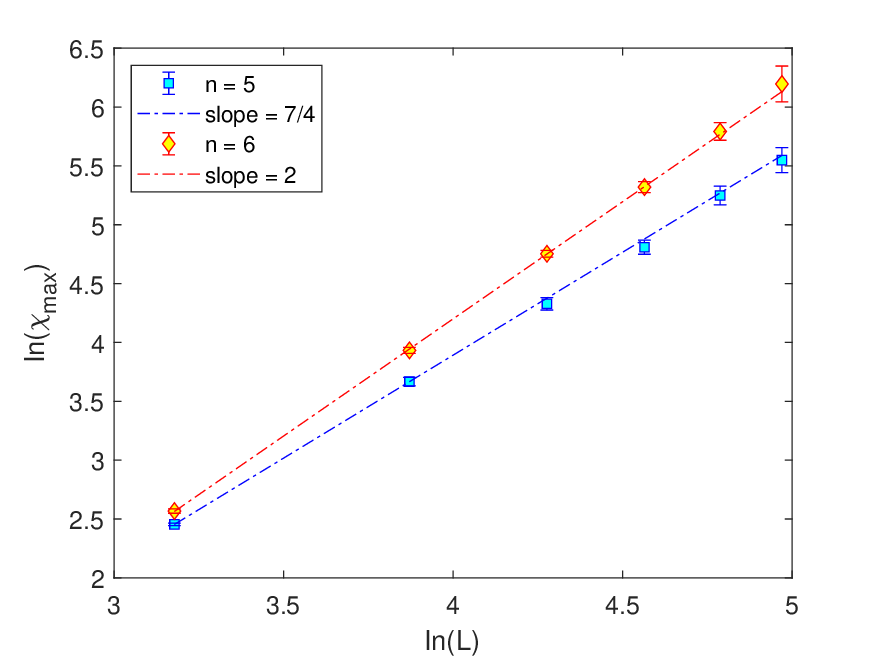}
\caption{FSS analysis of the magnetic susceptibility for the cases of $n=5$ and $n=6$.}
\label{fig:nk5-6_fss}
\end{figure}

Distinct nature of the phase transitions for $n=5$ and $n=6$ is further confirmed by a FSS analysis of the magnetic susceptibility, presented in Fig.~\ref{fig:nk5-6_fss}. For $n=5$ the slope of the linear fit of Eq.~(\ref{xi_FSS}) on a log-log scale gives the value of the critical exponent $\eta=1/4$ that corresponds to the BKT phase transition. On the other hand, for $n=6$ the slope equal to the lattice dimension tells us that the magnetic susceptibility scales with the lattice volume, as expected for a first-order phase transition.

\section{Discussion and Conclusion}
\label{sec:concusion}

\begin{figure}[t!]
\centering
\subfigure{\includegraphics[scale=0.5,clip]{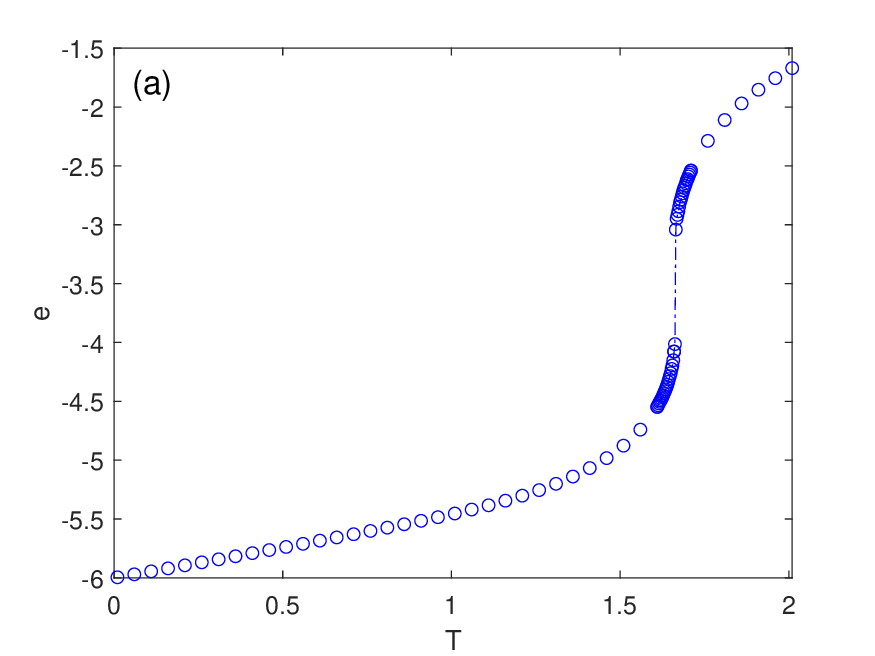}\label{fig:nk5_J02-1_ene}} 
\subfigure{\includegraphics[scale=0.5,clip]{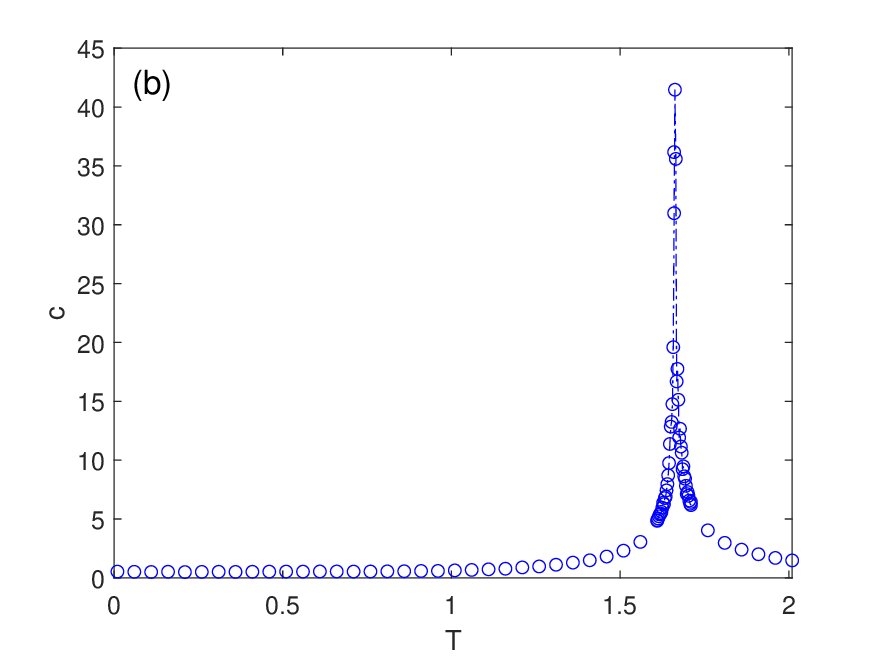}\label{fig:nk5_J02-1_c}} \\
\subfigure{\includegraphics[scale=0.5,clip]{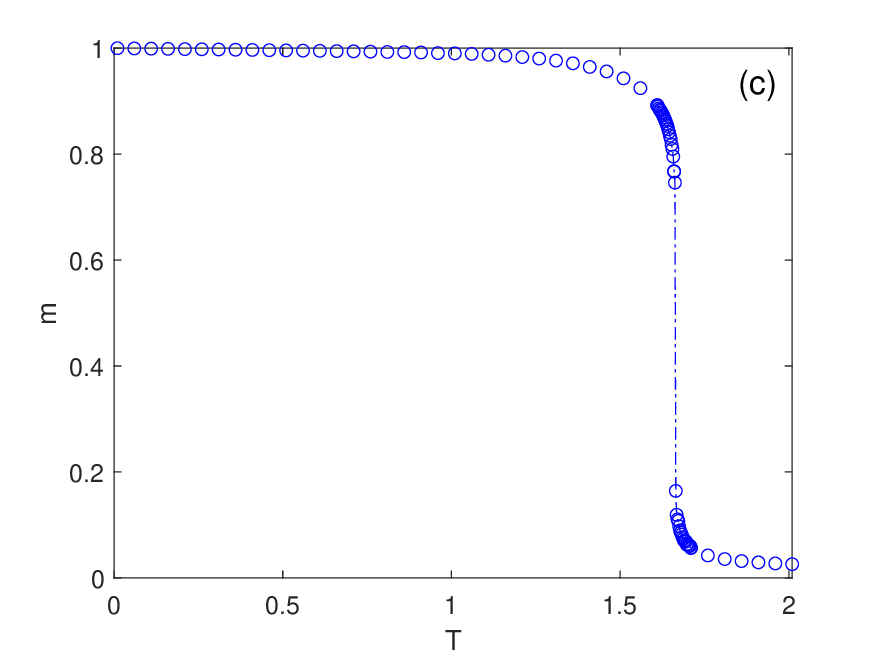}\label{fig:nk5_J02-1_m}}
\subfigure{\includegraphics[scale=0.5,clip]{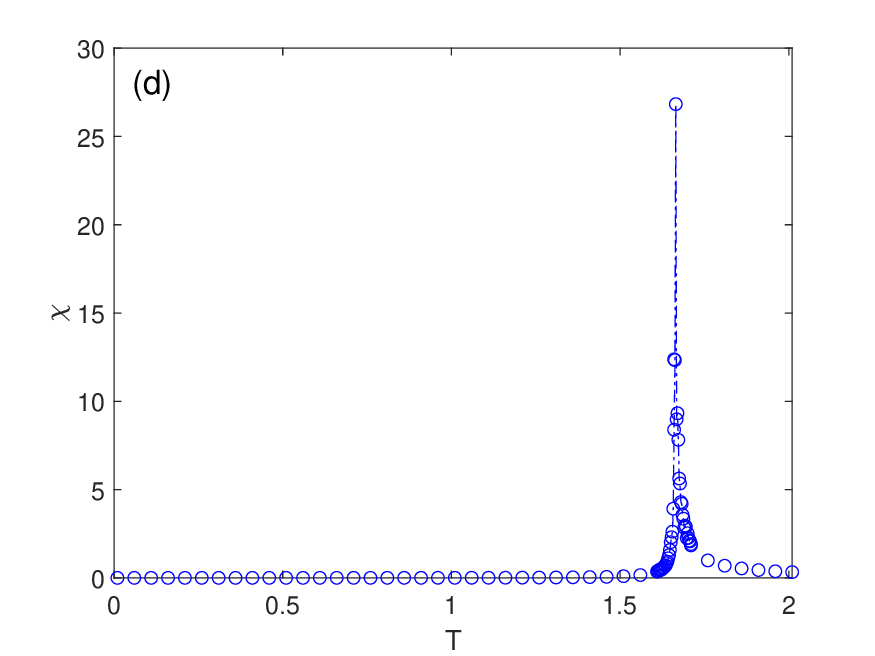}\label{fig:nk5_J02-1_chi}} \\
\subfigure{\includegraphics[scale=0.5,clip]{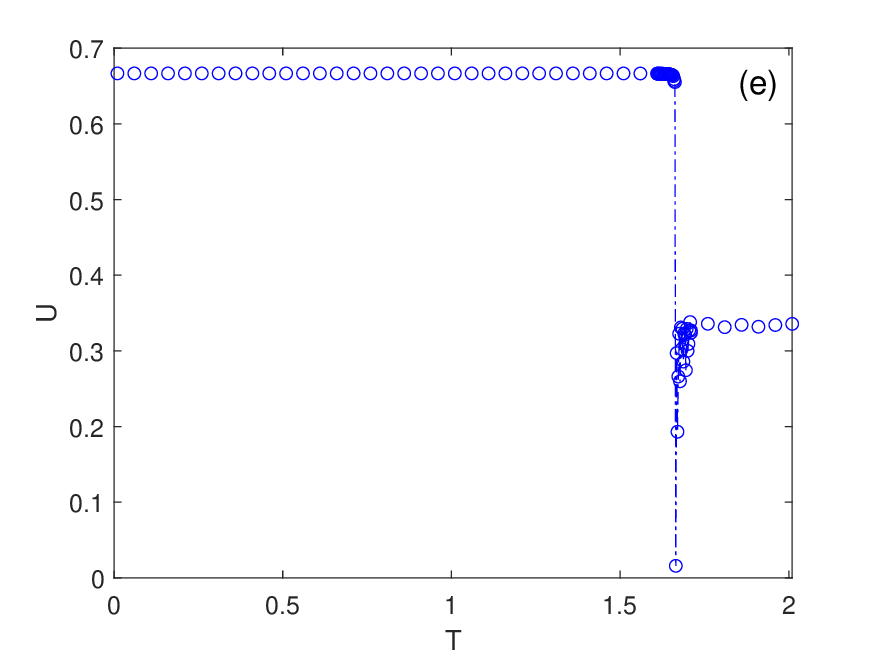}\label{fig:nk5_J02-1_u}}
\subfigure{\includegraphics[scale=0.5,clip]{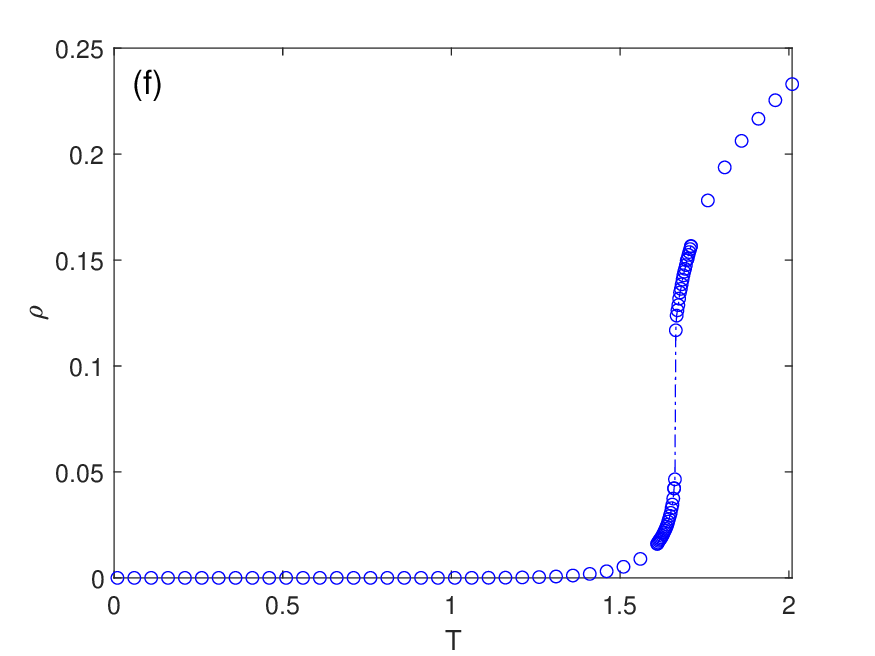}\label{fig:nk5_J02-1_rho}}
\caption{Temperature dependencies of (a) the internal energy, (b) the specific heat, (c) the magnetization, (d) the magnetic susceptibility, (e) the Binder cumulant, and (f) the vortex density, for $n=5$ with $J_1=0.2$, $J_2=0.4$, $J_3=0.6$, $J_4=0.8$, $J_5=1$, and $L=60$.}
\label{fig:x-T_nk5}
\end{figure}

From the above analysis it follows that in the considered generalized $XY$ model~(\ref{Hamiltonian1}), that involves $n$ terms with equal couplings, the minimal number of the terms necessary for the change of the nature of the transition from BKT to first order is $n_c=6$. One can ask whether a different setting of the coupling parameters can further reduce this number. The plots of the pairwise potential for a few selected settings of the coupling parameters, presented in Fig.~\ref{fig:well_n5}, suggest that the couplings $J_k$ with the gradually increasing values result in the potential well even narrower than for $J_k \equiv const$. We have checked if such conditions can reduce the value of $n_c$ by tentatively considering the model with $n=5$ terms and linearly increasing values of the coupling parameters: $J_1=0.2$, $J_2=0.4$, $J_3=0.6$, $J_4=0.8$, and $J_5=1$. The plots of the measured quantities as functions of temperature in Fig.~\ref{fig:x-T_nk5} indicate the presence of a single phase transition that bears some characteristics typical for a first-order transition, such as discontinuous behavior of the internal energy, magnetization, and vortex density, and spike-like peaks in the response functions. The fact that in Fig.~\ref{fig:nk5_J02-1_u} the Binder cumulant failed to reach negative values might be again attributed to an insufficiently small temperature step for the displayed extreme narrowness of the the dip after the transition point.

\begin{figure}[t!]
\centering
\subfigure{\includegraphics[scale=0.5,clip]{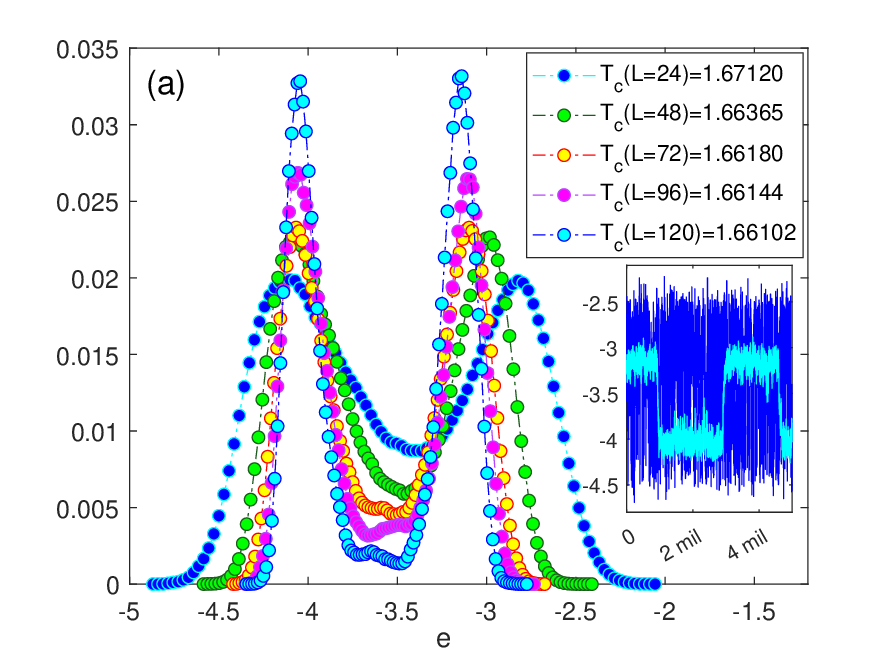}\label{fig:nk5_J02-1_ene_hist}}
\subfigure{\includegraphics[scale=0.5,clip]{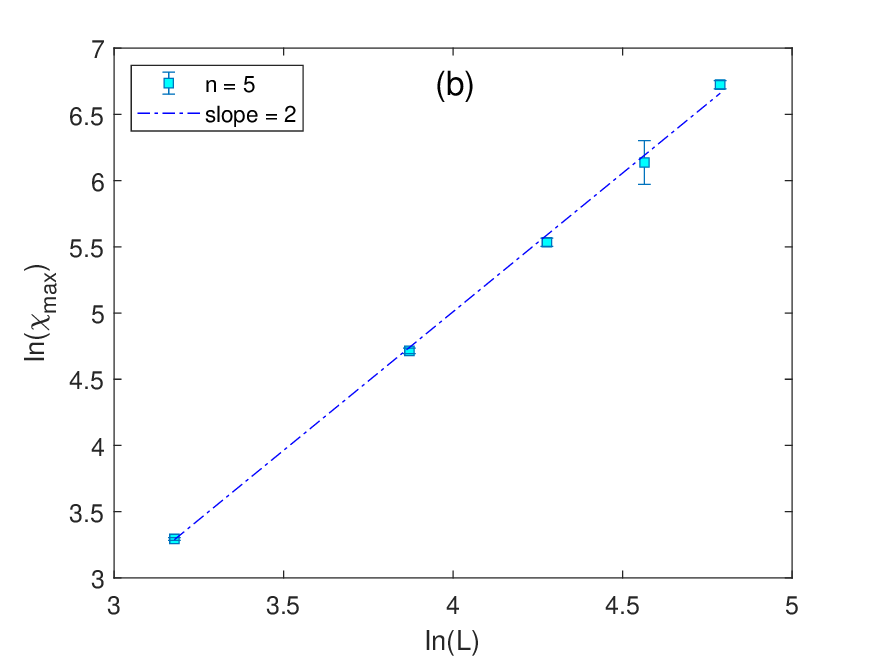}\label{fig:nk5_fss_J_incr}}
\caption{(a) Energy histograms at the (pseudo)transition temperatures for different lattice sizes and (b) FSS analysis of the magnetic susceptibility, for $n=5$ and $J_1=0.2$, $J_2=0.4$, $J_3=0.6$, $J_4=0.8$, and $J_5=1$.}
\label{fig:nk5_J02-1_ene_hist}
\end{figure}

Let us recall that the $n=5$ case for $J_k \equiv const$ only produced a pseudo-first-order transition. To confirm the true first-order nature of the transition for this case, in Fig.~\ref{fig:nk5_J02-1_ene_hist} we present a similar FSS analysis as done for the $J_k \equiv const$ case presented above. Both the analysis of the energy distributions and the magnetic susceptibility scaling provide rather clear evidence of the first-order transition. The first-order features are even stronger than for $n=6$ and $J_k \equiv const$ case. As shown in the inset of Fig.~\ref{fig:nk5_J02-1_ene_hist}, the tunneling times between the coexisting states are of the order of millions already for $L=120$ and thus inclusion of larger sizes would require much longer simulation times or more sophisticated simulation techniques.

To conclude, we have demonstrated that the setting of the coupling parameters $J_k$ with the equal or gradually increasing strength assigned to higher-order terms can cause that the first-order phase transition may appear in the present model with a relatively small number of terms. We have shown that it can appear in the model that includes $n=6$ terms with equal couplings but also in the model with only $n=5$ terms if their couplings gradually increase. One cannot rule out the possibility of the parameter setting that would result in the first-order transition occurring in the model with even smaller number of terms\footnote{We note that the $J_1-J_2$ model did not produce the first-order transition for any ratio $J_2/J_1$~\cite{lee85,kors85,carp89,geng09} and, therefore, the lower bound must be $n_c > 2$}. 

On the other hand, a too rapid increase of the couplings $J_k$ may lead to the splitting of the transition into two or possibly more separate transitions, instead of changing the nature of the single transition to first order. For example, in our tests (not shown) the exponentially increasing values $J_k=2^{^k}/2^{n}$, for $k=1,\hdots,n$ and $n=4$, led to the splitting of the transition into two separate transitions: the high-temperature transition appears to be of BKT but the nature of the low-temperature transition is unknown but most likely not first-order. However, the Hamiltonian~(\ref{Hamiltonian1}) can also be generalized to the forms that can show, in principle, even an arbitrary number of phase transitions as a function of temperature~\cite{zuko24}. Thus, we believe that it is worth to further explore different settings of the number of the higher-order terms and their interaction parameters as they can lead to a rich variety of the critical behavior.

\section*{CRediT authorship contribution statement}
\textbf{Milan \v{Z}ukovi\v{c}}: Conceptualization, Investigation, Methodology, Software, Data curation, Visualization, Formal analysis, Validation, Writing – original draft, Writing – review \& editing, Resources, Project administration.

\section*{Declaration of competing interest}
The author declares no competing financial interest.

\section*{Data availability}
No data was used for the research described in the article.

\section*{Acknowledgment}
This work was supported by the grants of the Slovak Research and Development Agency (Grant No. APVV-20-0150) and the Scientific Grant Agency of Ministry of Education of Slovak Republic (Grant No. 1/0695/23).

\bibliographystyle{elsarticle-num}
%\bibliography{multi}

\end{document}